\documentclass[a4paper,fleqn,usenatbib]{mnras}

\usepackage[T1]{fontenc}
\usepackage{ae,aecompl}

\usepackage{graphicx}    
\usepackage{amsmath}    
\usepackage{amssymb}    

\title[The Pan-STARRS1 map of halo substructures]{A Synoptic Map of Halo Substructures from the Pan-STARRS1 3$\upi$ Survey}

\author[E. J. Bernard et al.]{%
Edouard J. Bernard,$^{1,2}$\thanks{E-mail: ebernard@oca.eu (EJB)}
Annette M. N. Ferguson,$^{2}$
Edward F. Schlafly,$^{3}$
\newauthor
Nicolas F. Martin,$^{4,5}$
Hans-Walter Rix,$^{5}$
Eric F. Bell,$^{6}$
Douglas P. Finkbeiner,$^{7,8}$
\newauthor
Bertrand Goldman,$^{5}$
David Mart\'inez-Delgado,$^{9}$
Branimir Sesar,$^{5}$
\newauthor
Rosemary F. G. Wyse,$^{10,2}$
William S. Burgett,$^{11}$
Kenneth C. Chambers,$^{12}$
\newauthor
Peter W. Draper,$^{13}$
Klaus W. Hodapp,$^{12}$
Nicholas Kaiser,$^{12}$
Rolf-Peter Kudritzki,$^{12}$
\newauthor
Eugene A. Magnier,$^{12}$
Nigel Metcalfe,$^{13}$
Richard J. Wainscoat,$^{12}$
Christopher Waters$^{12}$
\\
$^{1}$Universit\'e C\^ote d'Azur, OCA, CNRS, Lagrange, France\\
$^{2}$SUPA, Institute for Astronomy, University of Edinburgh, Royal Observatory, Blackford Hill, Edinburgh EH9 3HJ, UK\\
$^{3}$Lawrence Berkeley National Lab, 1 Cyclotron Road, Berkeley, CA 94720, USA\\
$^{4}$Observatoire Astronomique de Strasbourg, Universit\'e de Strasbourg,
   CNRS, UMR 7550, 11 rue de l'Universit\'e, F-67000 Strasbourg, France \\
$^{5}$Max-Planck-Institut f\"ur Astronomie, K\"onigstuhl 17, D-69117
   Heidelberg, Germany \\
$^{6}$Department of Astronomy, University of Michigan, 500 Church St.,
   Ann Arbor, MI 48109, USA \\
$^{7}$Department of Physics, Harvard University, 17 Oxford Street, Cambridge, MA 02138, USA\\
$^{8}$Harvard-Smithsonian Center for Astrophysics, 60 Garden Street, Cambridge, MA 02138, USA\\
$^{9}$Astronomisches Rechen-Institut, Zentrum f\"ur Astronomie der
   Universit\"at Heidelberg, M\"onchhofstr. 12--14, D-69120 Heidelberg, Germany \\
$^{10}$Department of Physics and Astronomy, The Johns Hopkins University,
   3400 North Charles Street, Baltimore, MD 21218, USA \\
$^{11}$GMTO Corporation, 465 N.\ Halstead St., Suite 250, Pasadena, CA 91107, USA \\
$^{12}$Institute for Astronomy, University of Hawaii, 2680 Woodlawn Drive,
   Honolulu HI 96822, USA \\
$^{13}$Department of Physics, Durham University, South Road, Durham DH1 3LE, UK
}

\date{Accepted 2016 August 22. Received 2016 August 19; in original form 2016 July 19}

\pubyear{2016}

\begin{document}
\label{firstpage}
\pagerange{\pageref{firstpage}--\pageref{lastpage}}
\maketitle

\begin{abstract} 
We present a panoramic map of the entire Milky Way halo north of $\delta\sim-30\degr$ ($\sim$ 30,000 deg$^2$), constructed by applying the matched-filter technique to the Pan-STARRS1 $3\upi$ Survey dataset. Using single-epoch photometry reaching to $g\sim$22, we are sensitive to stellar substructures with heliocentric distances between 3.5 and $\sim$35~kpc. We recover almost all previously-reported streams in this volume and demonstrate that several of these are significantly more extended than earlier datasets have indicated.
In addition, we also report five new candidate stellar streams. One of these features appears significantly broader and more luminous than the others and is likely the remnant of a dwarf galaxy. The other four streams are consistent with a globular cluster origin, and three of these are rather short in projection ($\la10\degr$), suggesting that streams like Ophiuchus may not be that rare.
Finally, a significant number of more marginal substructures are also revealed by our analysis; many of these features can also be discerned in matched-filter maps produced by other authors from SDSS data, and hence they are very likely to be genuine. However, the extant $3\upi$ data is currently too shallow to determine their properties or produce convincing CMDs. The global view of the Milky Way provided by Pan-STARRS1 provides further evidence for the important role of both globular cluster disruption and dwarf galaxy accretion in building the Milky Way's stellar halo.
\end{abstract}



\section{Introduction}

One consequence of the hierarchical galaxy formation process predicted by cold dark matter cosmological models is that a significant fraction of the stellar mass in galaxies has been accreted. In disc galaxies like the Milky Way, stars that formed {\it ex situ} are overall a minority, but dominate the stellar halo \citep[e.g.][]{pil15}.
In these outer regions, where dynamical times are extremely long, the accreted material remains coherent for many billions of years \citep[e.g.][]{joh96}. Stellar streams are therefore powerful probes of the formation and evolution of galaxies:
in addition to providing direct evidence of past and ongoing accretion and disruption events, the observed properties of these substructures contain a wealth of information on both their progenitors and their host galaxy. For example, the stars from disrupted galaxies and globular clusters approximately follow, and therefore trace, the orbit of their progenitor, which provides an estimate of the mass and morphology of the potential enclosed within the orbit \citep[e.g.][]{kop10}.
The apparent width and velocity dispersion of globular cluster streams are strongly affected by density variations along their orbits, and can thus reveal the amount of clumpiness of the dark matter halo \citep[e.g.][]{iba02,nga16}. Finally, \citet{err15} have recently shown that the dark matter profile of dwarf spheroidal galaxies plays an important role in defining the sizes and internal dynamics of their tidal streams.

With the advent of wide-field photometric observations and surveys, many streams and substructures have been detected in the Milky Way \citep[see][and references therein; hereafter GC16]{gri16b} and in nearby galaxies \citep[e.g.][]{mal97,sha98,iba01,mar10,iba14,oka15,duc15,crn16}.
\defcitealias{gri16b}{GC16}
In the Galaxy, most of the known substructures have been discovered by searching for coherent stellar over-densities in the homogeneous, wide-field photometric catalogue provided by the {\it Sloan Digital Sky Survey} \citep[SDSS;][]{yor00}, although several streams have recently been found in other wide-field surveys \citep[e.g.][]{ber14a,kop14,mar14,bal16}. While some streams have clearly originated from the accretion of dwarf galaxies, about three quarters are consistent with globular cluster disruption according to \citetalias{gri16b}.
Since several teams have dedicated significant, independent efforts with the goal of detecting new substructures, one could expect that any stream within the detection limit of SDSS would have been found by now. However, like any survey, the SDSS catalogue suffers from artefacts, areas with shallower photometry due to e.g.\ weather conditions, and calibration issues revealing the observation patterns \citep[see e.g.][]{fin16}.

Here we present a systematic search for stellar substructures in the whole sky north of $\delta>-30\degr$ by taking advantage of the extensive coverage of the Pan-STARRS1 (PS1) 3$\upi$ Survey.
It significantly expands on the previous Milky Way substructure work that was carried out with an earlier data processing version of PS1 \citep{sla13,sla14,ber14b,mor16}.
The current processing version reaches to roughly the same depth as the SDSS but covers 30,000 deg$^2$ with homogeneous and well-calibrated photometry. The observational strategy and data reduction procedure are completely different from those of SDSS, thereby allowing a fully independent analysis. We first provide a summary of the substructures recovered in our analysis, including further extensions of known features, then present five new candidate streams, all but one of which lie within the SDSS footprint.

\section{The Pan-STARRS1 3$\upi$ Survey}

This work is based on the current internal data release of the PS1 3$\upi$ Survey (Processing Version 3; K.\ C.\ Chambers et al., in preparation), which covers the whole sky visible from Hawaii in five bands (Dec.$\ga-30\degr$; $g_{P1}r_{P1}i_{P1}z_{P1}y_{P1}$, hereafter $grizy$). The current depth of the catalogue, based on single-epoch photometry, reaches $g\sim$22 with a signal-to-noise ratio of 5. This corresponds to the old main-sequence turn-off (MSTO) magnitude of a stellar population at a heliocentric distance of $\sim$35~kpc, and thus probes a significant fraction of the Milky Way stellar halo.

The PS1 catalogue used in this paper is maintained by one of the authors (EFS) and stored in the Large Survey Database (LSD) format \citep{jur12}, which allows for a fast and efficient manipulation of very large catalogues ($>10^9$ objects). It contains both the point-spread function (PSF) and aperture photometry of each object, the difference between the two providing a convenient parameter to separate stars and background galaxies (see below).

\section{Data Analysis}

Our search for halo substructures in the 3$\upi$ dataset is based on the application of the matched-filter technique \citep{roc02}. In creating stellar density maps, this method gives higher weight to stars that are more likely to belong to an old and metal-poor (OMP; i.e. potentially accreted) component than to the main field population. The filter is built as the ratio of the Hess diagram of a OMP population to the Hess diagram of the field stars. The convolution is repeated by shifting the filter in magnitude to probe a range of heliocentric distances.

The matched-filter algorithm used here is based on the description of \citet{ode03}. It was written by one of us (EJB) as a MapReduce kernel for LSD to take advantage of the highly efficient, parallelized framework of LSD. This efficiency allowed us to carry out the matched-filtering over the whole PS1 footprint in a single run (i.e. $\sim$30\,000 deg$^2$), and to experiment with many different combinations of age, metallicity, heliocentric distance, and photometric bands in constructing the optimal filter.

\begin{figure*}
  \includegraphics[width=18.5cm]{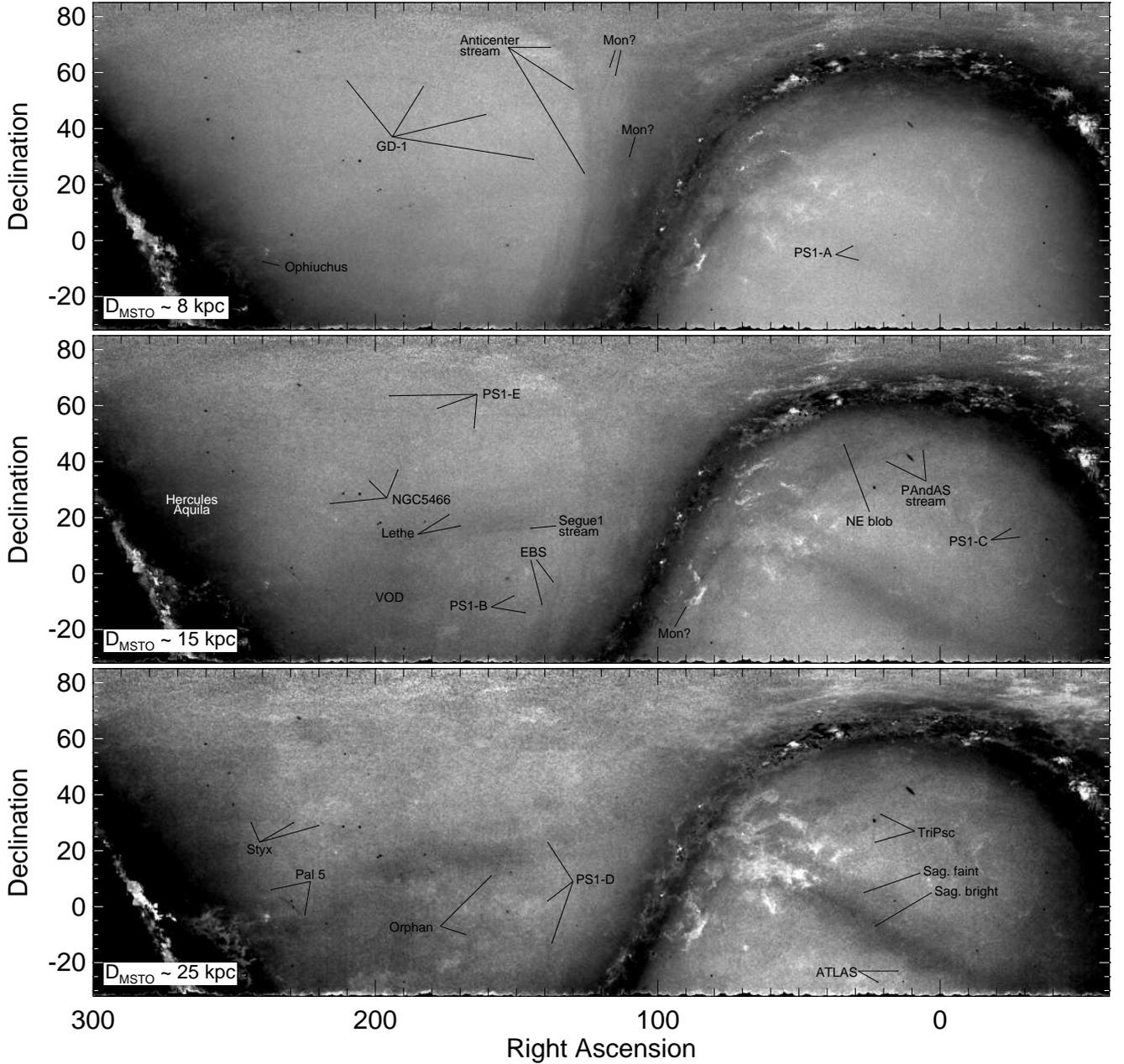}
  \caption{Matched-filtered stellar density maps of the whole PS1 footprint. The stretch is logarithmic, with darker areas indicating higher surface densities; a few black patches are due to missing data. The top, middle, and bottom panels correspond broadly to heliocentric distances of 8.5, 15, and 25~kpc. The main substructures are labeled and discussed in Section~\ref{sec:known}; new candidate streams (labelled PS1-A to E) are presented in Section~\ref{sec:new}.}
  \label{fig:maps}
\end{figure*}

While many studies to date have used the CMD of the globular cluster M13 \citep[NGC\,6205; e.g.][]{gri09,bon12} to build the filter, we have chosen to use synthetic colour-magnitude diagrams (CMDs). This implementation has several advantages: (i) we can generate arbitrarily well populated CMDs at all magnitudes; (ii) these CMDs are not contaminated by field objects; (iii) we can simulate photometric uncertainties adequately for all distances probed without having first to correct for the actual uncertainties of the M13 observations; and (iv) we can repeat the convolution with different combinations of ages and metallicities. The synthetic CMDs were generated from PARSEC isochrones \citep{bre12} in the PS1 bands\footnote{downloaded from version 2.8 of http://stev.oapd.inaf.it/cgi-bin/cmd} following the luminosity function provided in the isochrone files, and corrected for completeness as a function of magnitude as measured in the representative region described below.
In contrast, the field Hess diagram was produced empirically by selecting all the stellar objects within the region defined by 215$\degr$ < RA < 245$\degr$ and 15$\degr$ < Dec. < 60$\degr$, corresponding to
an area of $10^3$ deg$^2$ containing $\sim4.9\times10^6$ objects. This region was chosen because it does not contain any known Local Group dwarf galaxy or globular cluster. In addition, as it encompasses a wide range of Galactic latitudes, it is representative of the field population over most of the sky where substructures are likely to be detected with this method.

\begin{figure*}
  \includegraphics[height=6.2cm]{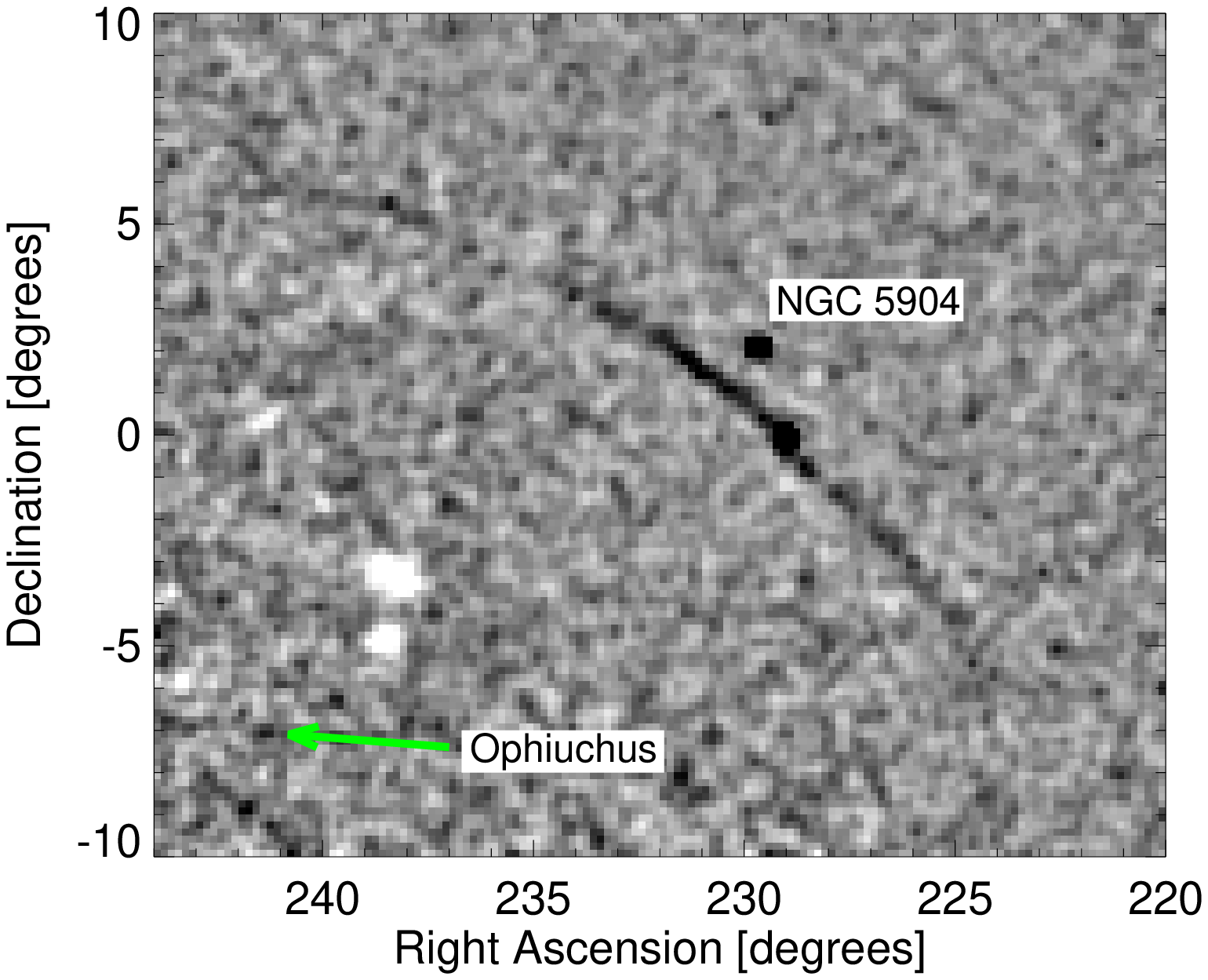}
  \includegraphics[height=6.2cm]{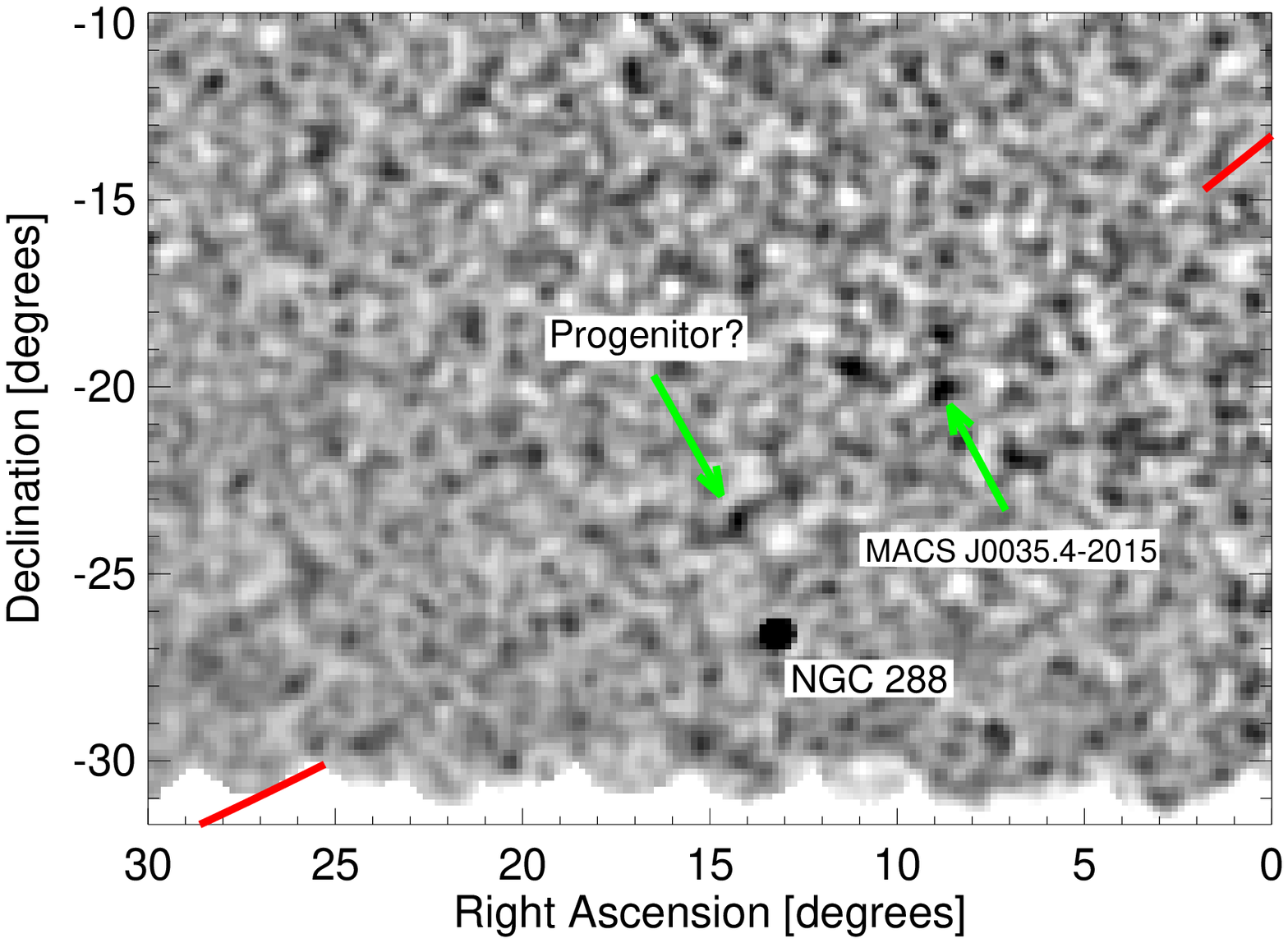}
  \caption{Close-up view of the area around Pal\,5 (left) and the ATLAS stream (right). To remove large-scale background variations, the maps were smoothed with a Gaussian kernel of width 2$\degr$ and subtracted from the original maps.
  Some fore- and background features are labeled. The red line in the right panel shows the best fitting great circle containing the stream, with north pole at ($\alpha$, $\delta$)=(74\fdg599,48\fdg365).}
  \label{fig:pal5}
\end{figure*}

Only stellar-like objects (i.e.\ defined as $|r_{\rm PSF}-r_{\rm aperture}| \leq0.2$) with photometric uncertainties below 0.2~mag in either ($g$ and $r$) or ($g$ and $i$) were taken into account. They were then corrected for foreground reddening by interpolating the extinction at the position of each source using the \citet{sch14} dust maps with the extinction coefficients of \citet{sch11}.
The matched-filtering was carried out in 26 heliocentric distance slices from 3.5 to 35~kpc (i.e. distance moduli from 12.7 to 17.7 separated by 0.2~mag) in both $(g-r, g)$ and $(g-i, g)$ filter combinations. After visual examination of the different slices, we decided to coadd slices 8 to 13, 14 to 19, and 20 to 25, to produce three maps for each filter combination, corresponding broadly to heliocentric distances of $\sim$8.5, 15, and 25~kpc. Finally, we corrected for the different pixel area over the sky for better contrast at high declination, and averaged the $(g-r, g)$ and $(g-i, g)$ maps together to produce the maps shown in Fig.~\ref{fig:maps}. The pixel scale is 5$\arcmin$, smoothed by a Gaussian kernel with a full-width at half-maximum (FWHM) of 3 pixels for the 8.5 and 15~kpc maps, and 4 pixels for the 25~kpc map. We found this angular resolution to be the best compromise between revealing the broad, diffuse substructures and smoothing out the cold streams. Note, however, that some features (e.g.\ PS1-E, see below) are much more prominent when using a broader Gaussian filter.

We repeated the filtering for different combinations of age (8, 10, 12, and 13.5 Gyrs) and metallicity ([Fe/H] = -2.2, -1.9, -1.5, -1.0), but found that (12, -1.5) produced the best overall contrast. The FITS file containing the 26 distance slices from 3.5 to 35~kpc based on that age/metallicity combination, and a second file with the three coadded, unsmoothed maps used to create Fig.~\ref{fig:maps}, are made available online\footnote{http://dx.doi.org/10.5281/zenodo.60518}.

\section{Recovery of Known Substructures}\label{sec:known}

The majority of known stellar streams and substructures have been discovered thanks to the \emph{Sloan Digital Sky Survey} \citep[SDSS;][]{yor00}, which observed roughly 14,555 square degrees of sky at a comparable depth to the PS1 3$\upi$ survey. This area is completely contained with the PS1 footprint and
hence our ability to recover these features provides a check on the photometric accuracy and purity of the current catalogue. As a reference database, we use the recent compilation of Milky Way halo streams presented by \citetalias{gri16b} and complement this with additional features which have been subsequently discovered by \citet{bal16} and \citet{bel16} using data from the \emph{Dark Energy Survey} \citep[DES;][]{des05}.

Our maps clearly reveal all the prominent structures that have been reported, but also most of the fainter detections. In particular, we recover the bright and faint streams from the Sagittarius dwarf galaxy \citep{iba94,bel06b}, the tails of Palomar\,5 \citep{ode01}, the Orphan stream \citep{bel06b,gri06c}, GD-1 \citep{gri06b}, the Anticentre stream and the Eastern Banded Structure \citep[EBS;][]{gri06d}, the ATLAS stream \citep{kop14}, the Pisces--Triangulum stream \citep{bon12}, the Ophiuchus stream \citep{ber14a}, the PAndAS MW stream and the PAndAS northeast blob \citep{mar14}, the tidal feature around Segue\,1 \citep{nie09}, the Hercules--Aquila cloud \citep{bel07a}, and the Virgo Overdensity \citep[VOD][]{ive00}.  These features are labelled in the map of their corresponding distance slice as shown in Fig.~\ref{fig:maps}.
A close examination of the maps in FITS format also reveals the tails of NGC\,5466 \citep{bel06a}, as well as portions of Lethe and Styx \citep{gri09}, and Hyllus and Hermus \citep{gri14a}.

\begin{figure}
  \includegraphics[width=\columnwidth]{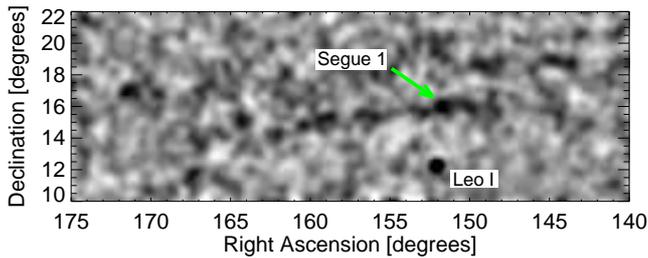}
  \caption{Same as Figure~\ref{fig:pal5} for the stream in the vicinity of Segue\,1, which extends from R.A. of $\sim$144$\degr$ to 168$\degr$.}
  \label{fig:seg1}
\end{figure}

On the other hand, there are a few features listed in the \citetalias{gri16b} compilation that we do not recover -- namely, the Virgo Stellar stream \citep[VSS;][]{duf06}, Acheron and Cocytos \citep{gri09}, the Cetus Polar stream \citep[CPS;][]{new09}, the Pisces overdensity \citep[POD;][]{ses07}, and TriAnd \citep{maj04,mar07}.
However, in most of these cases, the absence of these features in our maps is understandable: we note that both the VSS and CPS are broad and diffuse streams that have not yet been observed in matched-filter or MSTO maps, the POD lies more than twice as far away as the reach of our photometry, and Acheron and Cocytos are located in a dense stellar region near the bulge, which complicates their detection. While we do not observe any well-defined overdensities that could correspond to TriAnd, we note that the middle and bottom panels of Fig.~\ref{fig:maps} do reveal low-level diffuse structure in the vicinity of the Andromeda galaxy. In summary, we believe that there are good reasons why these specific features will not be visible in PS1 and emphasize the excellent overall recovery rate we have achieved for known streams over this large swath of the sky.

Four stellar streams have been discovered to date that lie outside the PS1 footprint (at $\delta<-30\degr$), namely Alpheus \citep{gri13}, streams S2 and S3 \citep{bel16}, and Phoenix \citep{bal16}. It has recently been argued that the latter may be an extension of the Hermus stream \citep{gri14a,gri16a}. Based on their known trajectories, we conducted a search for possible extensions of all these streams in the PS1 footprint, but did not find any significant overdensities. That said, both S2 and S3 have heliocentric distances larger than 50~kpc and are therefore also beyond the volume sampled here.
 
In several cases, the PS1 data provide new constraints on the properties and/or spatial extension of known streams. In the following subsections, we discuss a number of the specific features. Note that Sagittarius and Monoceros have already been discussed extensively in the context of the PS1 data \citep{sla13,sla14,her16,mor16}.

\subsection{The tidal tails of Palomar 5}

The Galactic globular cluster Palomar\,5 (Pal\,5) harbors the most prominent tidal tails among the known Milky Way clusters, and naturally these have been extensively studied.
The tails were first discovered by \citet{ode01} using small area SDSS commissioning data, but subsequent SDSS data releases with expanding coverage allowed them to be traced far further. The currently known length of the Pal\,5 stream is $\sim$$22\degr$ on the sky \citep{gri06e}, though it is truncated in the south by the edge of the SDSS survey at $\delta=-2.5\degr$.

\begin{figure*}
  \includegraphics[height=3.85cm]{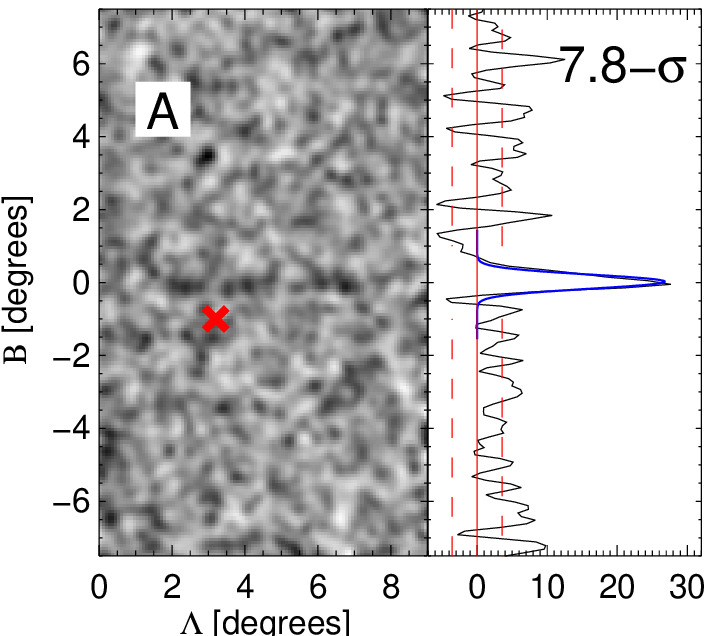}
  \includegraphics[height=3.85cm]{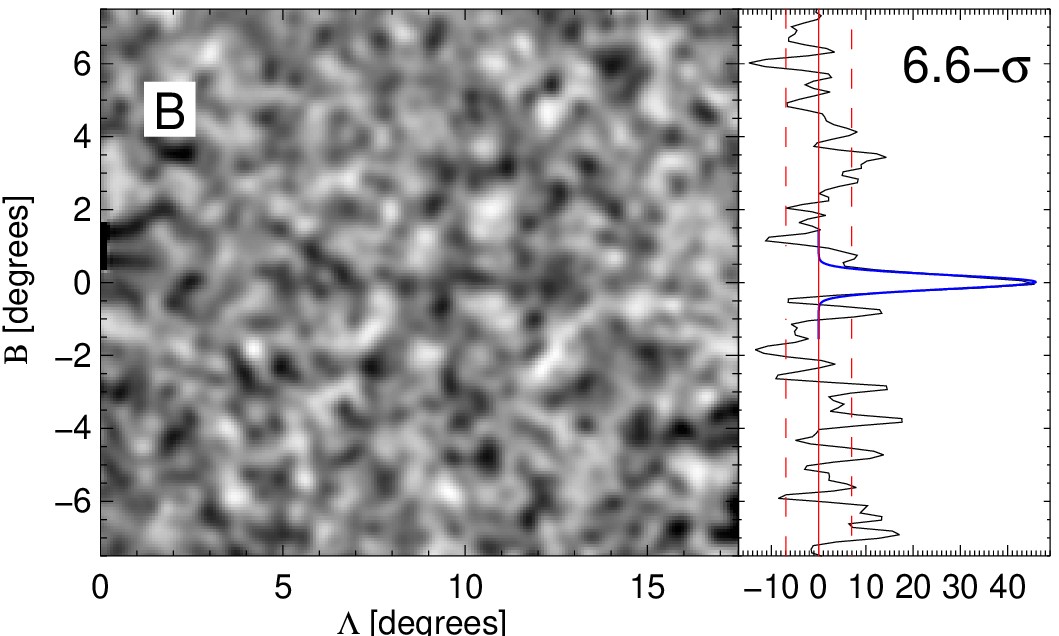}
  \includegraphics[height=3.85cm]{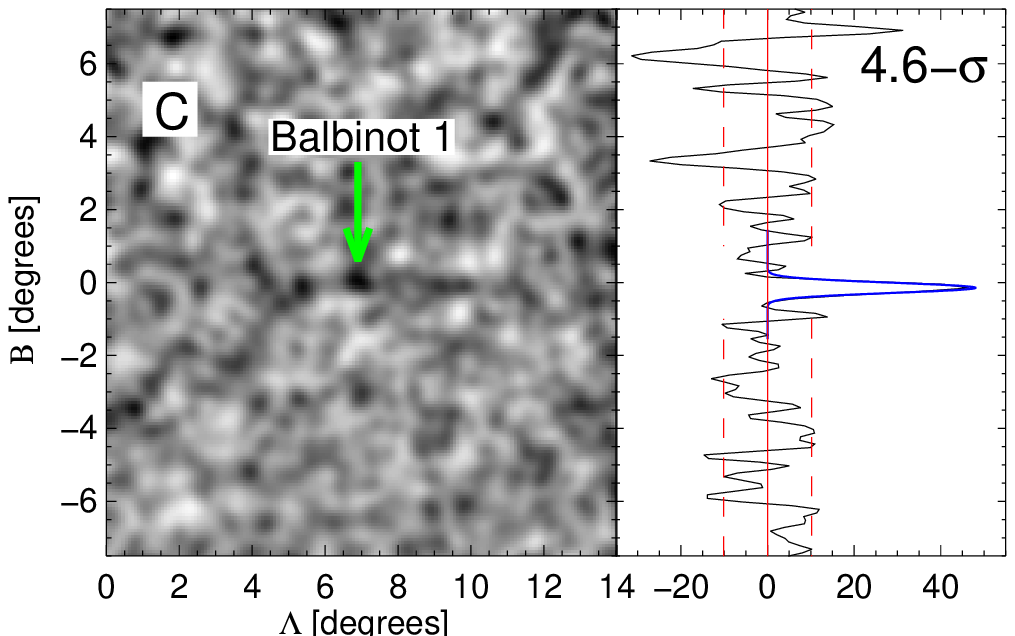}
  \includegraphics[height=4.8cm]{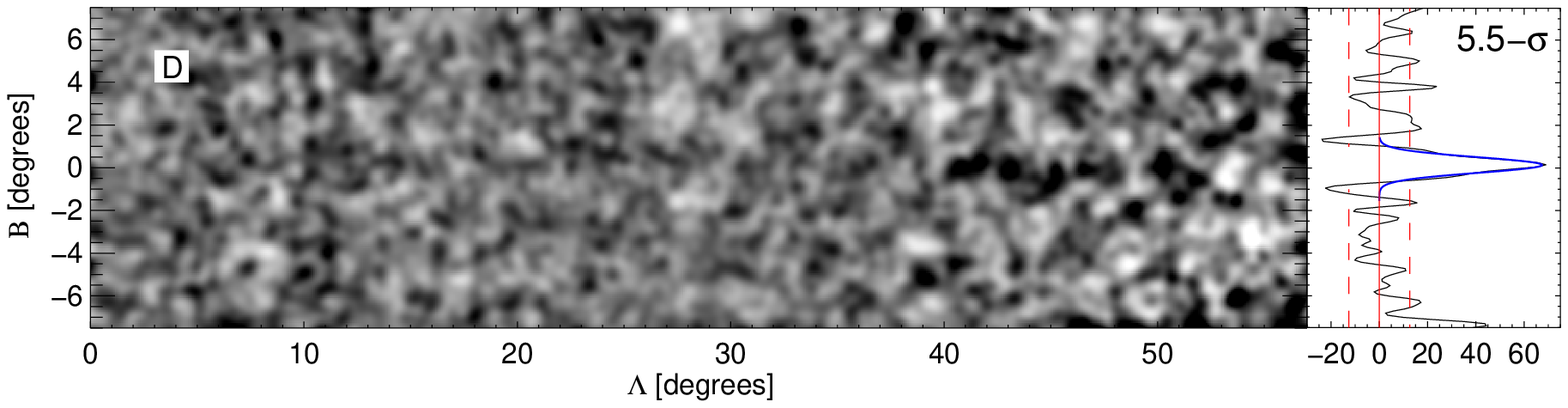}
  \includegraphics[height=3.85cm]{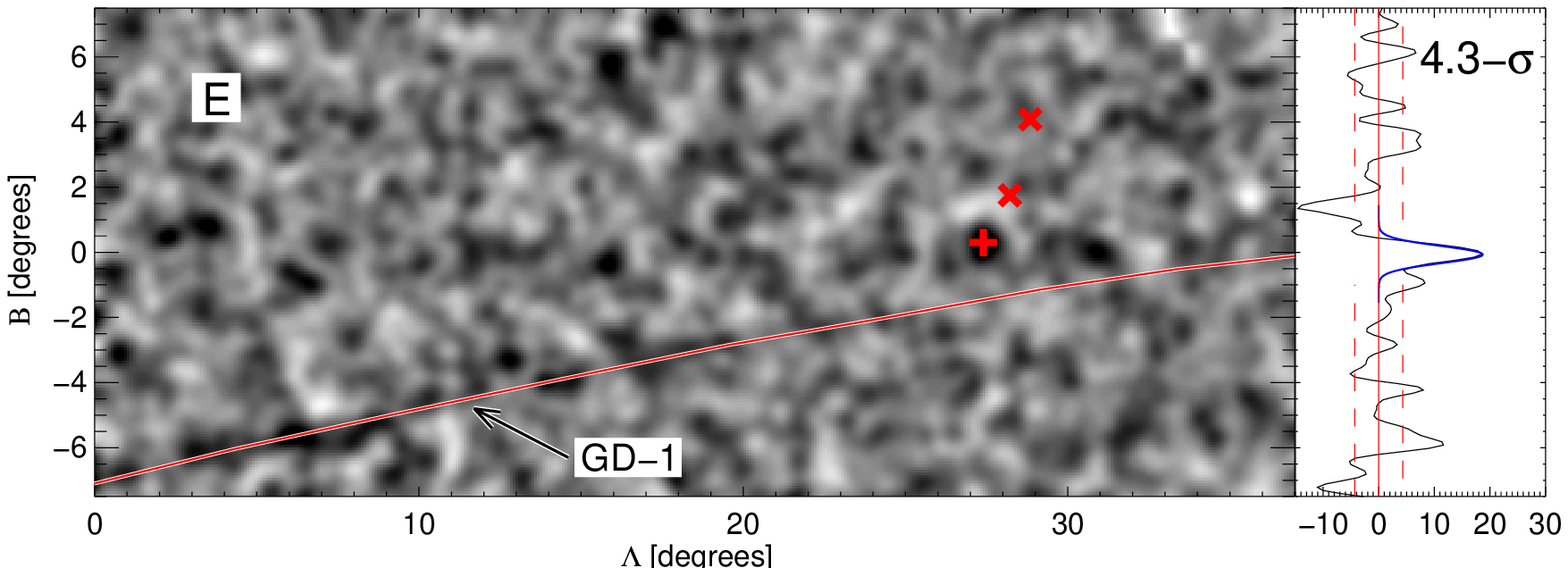}
  \caption{Close-up view of the new stream candidates. For each stream, we show
  the matched-filtered map in the coordinate system of the stream ({\it left}) and
  the average stellar density projected onto the $B$ axis in arbitrary units ({\it right, see text}).
  In the maps, which are the coaddition of the four distance slices in which the stream
  signal is strongest, the streams lie at $B\sim$~0 with the right hand side being the southern-most point.
  In the cross-section panels, the dashed lines show the 1-$\sigma$ dispersion
  of the background, while the blue line is a Gaussian fit to the stream profile.
  The significance of detection is also indicated.
  {\it PS1-A:} The red cross in the left panel shows the Galactic globular
  cluster Whiting\,1, which is unrelated to the stream.
  {\it PS1-C:} Even after masking a 1 deg$^2$ area around Balbinot\,1 -- at $(\Lambda,\rm{B})\sim (7,0)$ -- the peak of PS1-C still reaches 4.1-$\sigma$.
  {\it PS1- D:} The peak of the profile is at 4.8-$\sigma$, or 3.4-$\sigma$ for $\Lambda < 40\degr$.
  {\it PS1-E:} The peak of the profile is at 4.3-$\sigma$ when masking a 1 deg$^2$ area around the overdensity marked with a plus sign at ($\Lambda$,$B$)$\sim$(27.5,0.4). The cross symbols represent Willman\,1 \citep[lower;][]{wil05a} and Ursa Major\,I \citep[upper;][]{wil05b}. The narrow stream highlighted in red is GD-1.}
  \label{fig:newst}
\end{figure*}

While there have been numerous observational efforts to obtain deeper photometric data \citep[e.g.][]{iba16}, as well as spectroscopic measurements \citep[e.g.][]{ode09}, to constrain the properties and possible orbits of the stream, these have focused on the portion of the stream which lies inside the SDSS footprint. With a spatial coverage extending about 30$\degr$ further to the south, PS1 allows us to search for extensions of the stream beyond this area. From our maps (see Figure~\ref{fig:pal5}), we are able to trace the Pal\,5 stream further south to $\delta=-6\degr$, where it appears to end abruptly. This is not an observational bias due to e.g.\ a distance gradient along the stream, since in that case we would expect to see the stream appear in nearer/more-distant slices.
This means that the leading ($\sim8\degr$) and trailing ($\sim15\degr$) tails of Pal 5 are of distinctly different traceable angular extent.

\subsection{The ATLAS stream}

The ATLAS stream was recently discovered by \citet{kop14} from early data of the ATLAS survey \citep{sha15}. They were able to trace it over 12$\degr$, from $\delta\sim-32\degr$ to the edge of a gap in the spatial coverage at the time at $\delta=-25\degr$; there appeared to be no continuation of this feature in the narrow ATLAS stripe covering $-13.5\degr<\delta<-10\degr$. In our maps -- zoom-in map is shown in Figure~\ref{fig:pal5} -- we find that the stream extends significantly further north to about $\delta=-15\degr$. It therefore appears that the combination of the ATLAS and PS1 surveys covers the entire $\sim28\degr$ length of this stream.

\begin{figure*}
  \includegraphics[height=6cm]{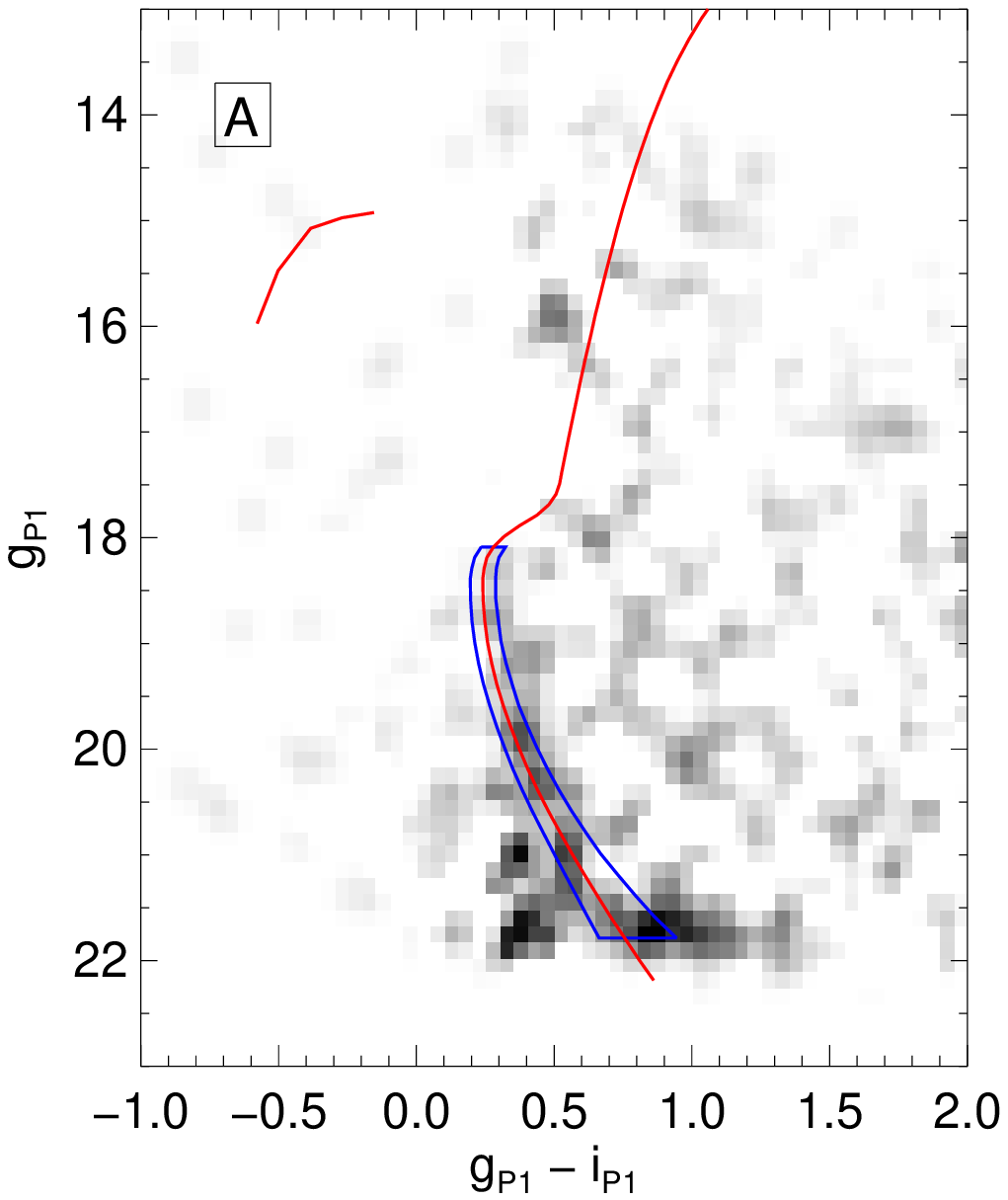}
  \includegraphics[height=6cm]{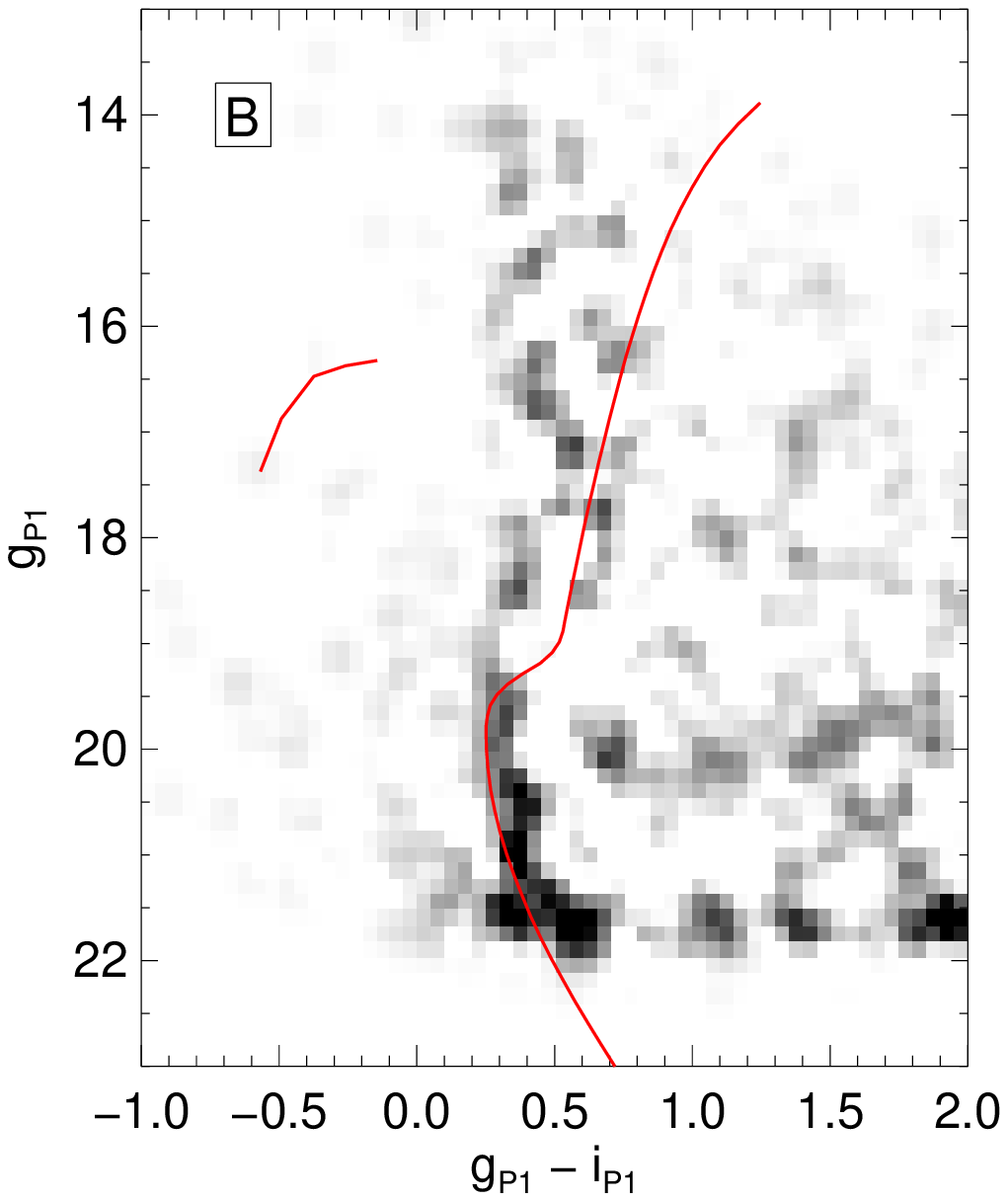}
  \includegraphics[height=6cm]{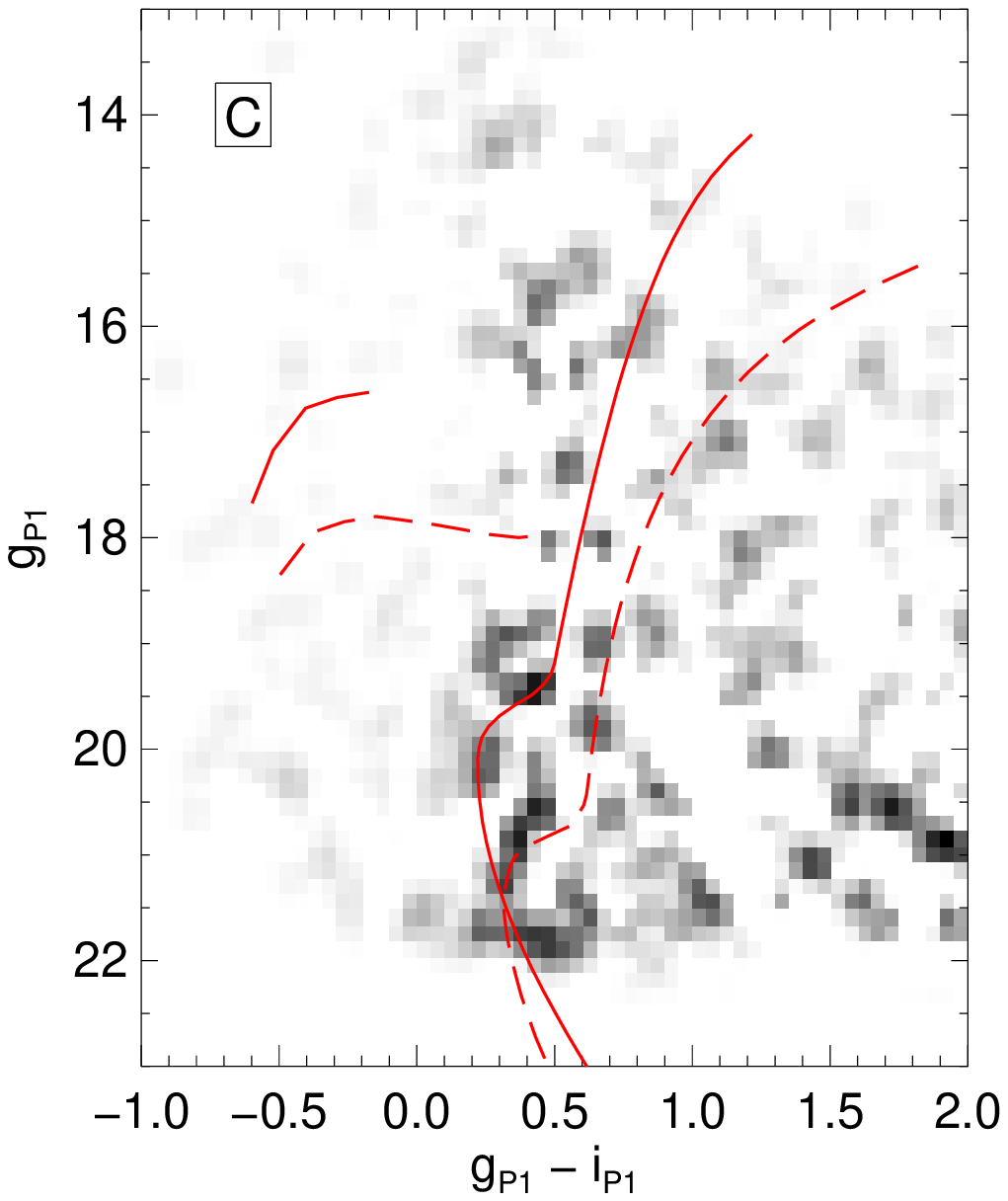}
  \includegraphics[height=6cm]{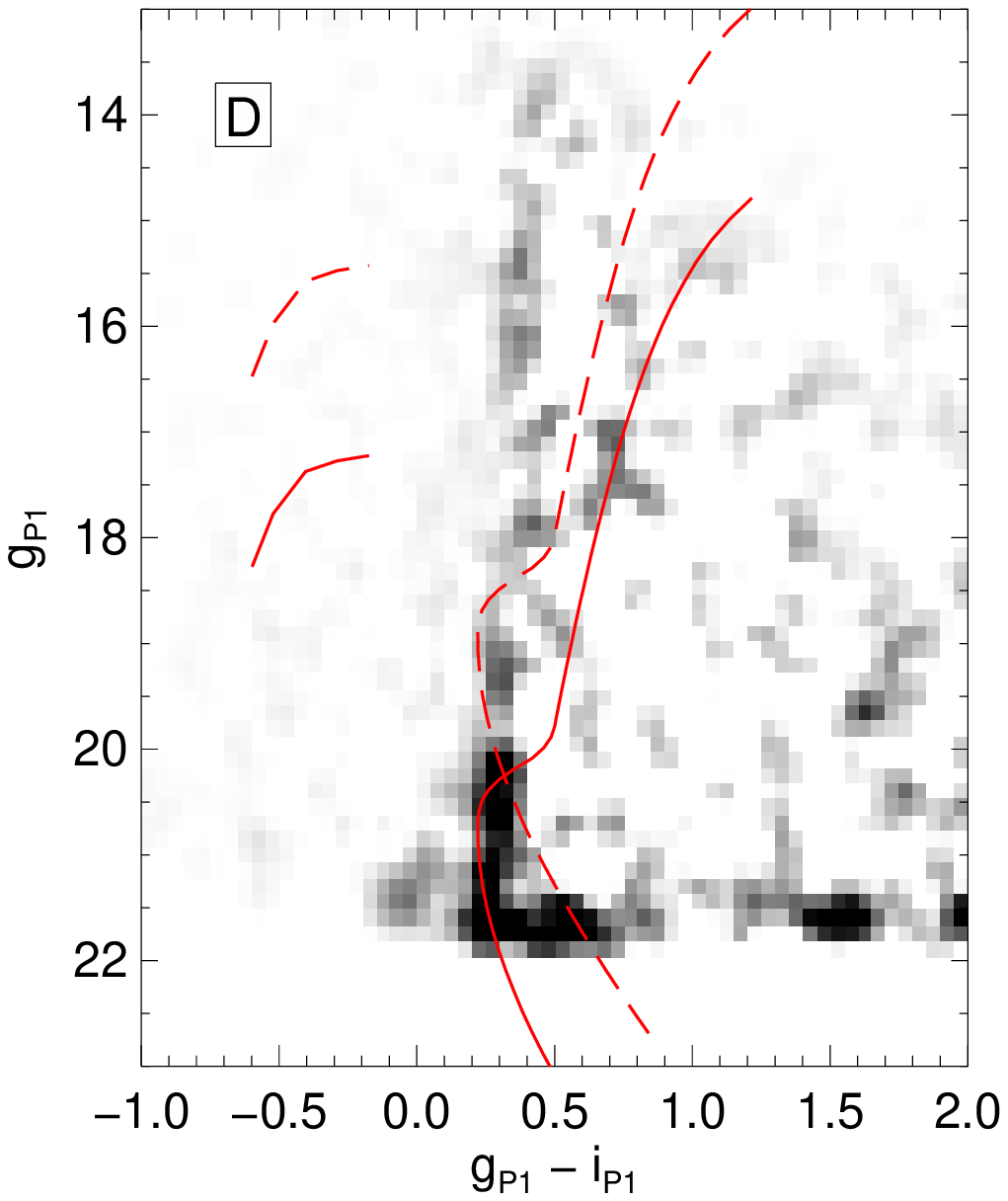}
  \includegraphics[height=6cm]{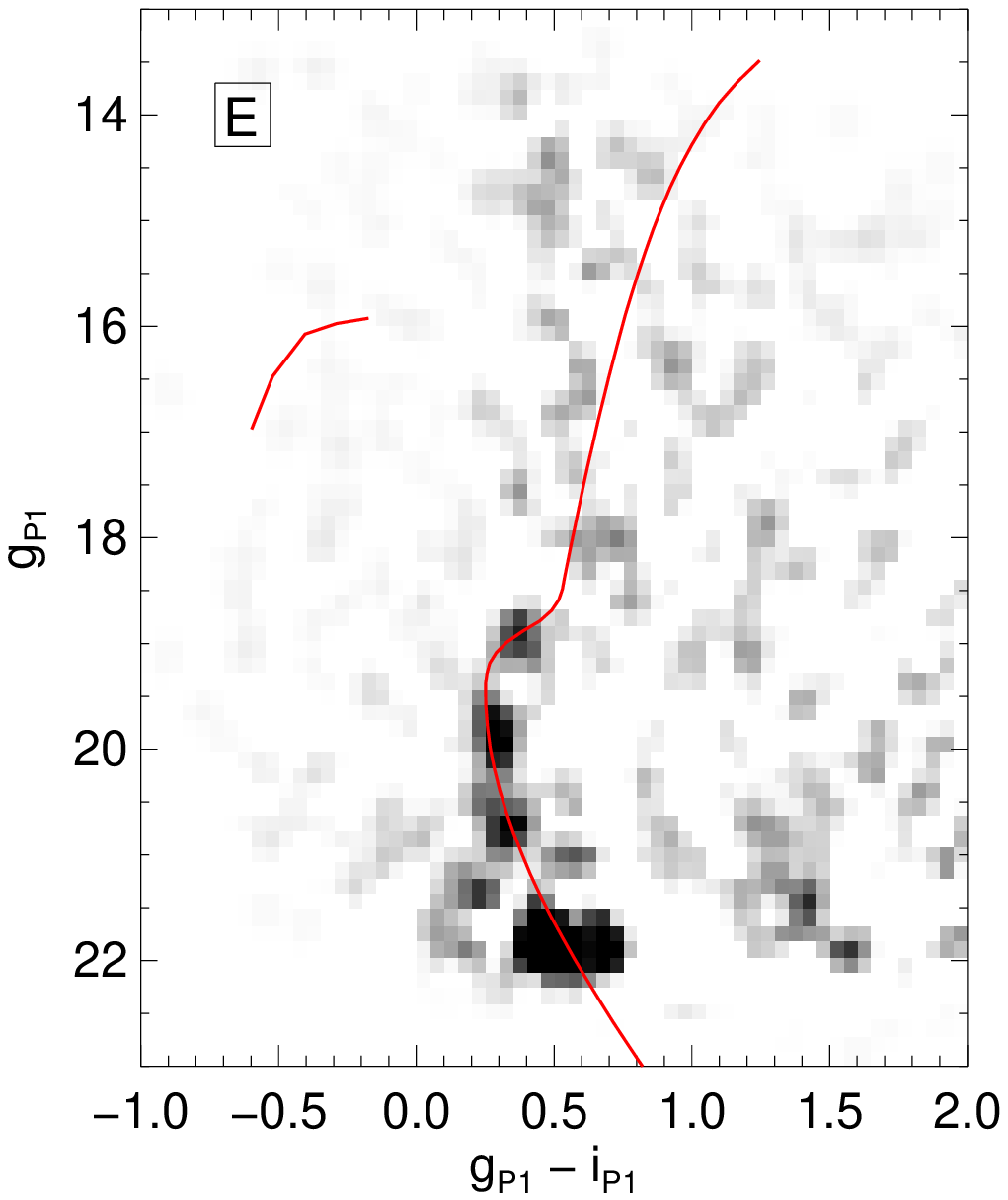}
  \caption{Extinction-corrected CMDs of the new stream candidates. The red lines are globular cluster fiducials from \citet{ber14b}, shifted to the distance obtained from the matched-filtered slices where the signal of the streams was stronger. The dashed fiducials in panels C and D correspond to other substructures in the line of sight, complicating the CMDs: Balbinot\,1 and the EBS, respectively. Balbinot\,1 is represented by the fiducial of M3, while the fiducial of M92 is used for the other features. In panel A, the blue polygon represents the selection box used to select stream member stars for the significance and width measurements; similar boxes have been used for each stream, adapted to their distance and photometric uncertainties.}
  \label{fig:cmds}
\end{figure*}

The narrow width of the stream ($\sim 0.25$~deg) combined with the metal-poor nature of its stellar population led to \citet{kop14} to conclude that the progenitor was a globular cluster. A compact overdensity is visible in Fig.~\ref{fig:maps} along the stream at ($\alpha,\delta$) $\sim$ (0$^h$35$^m$, $-20\degr$05$^m$). However, inspection of the stacked images reveals a background galaxy cluster (MACS J0035.4-2015) is most likely responsible for this enhancement. Aside from this, there is also a significant, broader overdensity at ($\alpha,\delta$) $\sim$ (0$^h$57$^m$, $-23\degr$29$^m$), which roughly coincides with the centre of the stream. Interestingly, the matched-filter maps suggest this feature has an -- admittedly low significance -- S-shaped morphology, as would be expected for a disrupting globular cluster. There is no obvious stellar concentration visible at this position in the stacked images and hence deeper imaging will be necessary to confirm its presence.

Based on simple orbit modelling, \citet{kop14} suggested that a possible progenitor of the ATLAS stream might be the sparse halo globular cluster Pyxis. Pyxis lies at ($\alpha,\delta$) $\sim$ (9$^h$08$^m$, $-37\degr$13$^m$) and hence outwith the PS1 footprint. Although we cannot directly test this association with our data, our detections of the stream at more northern latitudes will enable significantly improved constraints on possible progenitor orbits.

\subsection{The stream in the vicinity of Segue\,1}

Based on an analysis of SDSS data, \citet{nie09} report tidal tails extending to $\sim1\degr$ both eastward and southwest of Segue\,1. A matched-filter analysis of the same data by \citet{gri14b} suggests that this field contains a narrow stream extending over at least 25$\degr$, but lying a few kpc closer than Segue\,1.  Follow-up spectroscopic analysis of this region \citep{geh09,nor10,sim11} led to the detection of a cold component with $v_{helio}\sim300$~km\,$s^{-1}$ having stellar population and heliocentric distance roughly comparable to those of Segue\,1, but significantly offset in radial velocity. Unfortunately, the spectroscopic observations only cover a small area around Segue\,1, so it is not clear yet how the tidal extensions, the 25$\degr$ stream and the 300 km\,s$^{-1}$ velocity component are connected. The region around Segue\,1 is further complicated by the fact that the Sagittarius stream lies in the background and may possess a component with such a radial velocity.

Inspection of our maps recovers a $\sim24\degr$ stream crossing Segue\,1 (see Figure~\ref{fig:seg1}).
Consistent with \citet{gri14b}, we find this feature lies several kpc closer than Segue\,1 which is at a heliocentric distance of $\sim$23~kpc. There appears to be a distance gradient along the stream with the eastern end closer to the Sun than the western end ($\sim$14kpc vs. $\sim$19kpc). This gradient is opposite to that of the Sagittarius Stream in the same longitude range \citep[e.g.][]{bel06b}, which further strengthens the case of the stream being separate from both Segue\,1 and Sagittarius \citep[e.g.][]{fre13}. Confirmation of the association with the 300 km\,s$^{-1}$ population will require spectroscopic observations of stream members several degrees away from Segue\,1.

\subsection{Monoceros and the anticentre substructures}

The Milky Way disc in the direction of the anticentre contains a number of features whose nature and origin are still hotly debated \citep[see][for a detailed description and historical review]{sla14,mor16}. This low-latitude substructure is often referred to as the Monoceros Ring, after the constellation in which the first evidence was discovered. It forms a large and complex stellar enhancement in the outer disc, mainly confined between 14-18 kpc from the Galactic centre, and extending from $120\degr< l < 240\degr$ and $-30\degr < b < +40\degr$. It is most visible in the top panel of Figure~\ref{fig:maps} between RA of $\sim$110 and $-20\degr$ and on either sides of the disc. It has a rather sharp edge, as might be expected from a flaring of the outer disc.
Recent work has shown that these low-latitude features could primarily be the result of disc oscillations \citep{xu15}, possibly as a consequence of a low-mass satellite fly-by \citep[e.g.][]{gom16}.

In our maps, several new, well-defined substructures in the anticentre region can be identified and we label these `Mon?', as it is unclear if and how they are related to Monoceros. For example, the two parallel marks labeled `Mon?' in the top panel of Figure~\ref{fig:maps} lie within 2$\degr$ of the EBS orbit projections from \citet{gri08}, and may therefore be an extension of this feature.

\begin{table*}
\centering
\caption{Summary of the Properties of the Newly-Discovered Candidate Streams.}
\label{tab:prop}
\begin{tabular}{lccccc}
\hline
 Parameter             & PS1-A                                        & PS1-B                                        & PS1-C                                        & PS1-D                                         & PS1-E \\
\hline
 R.A.\ (J2000.0)       & \ \ 01:57:02                                 & \ \ 09:53:15                                 & \ \ 22:10:28                                 & \ \ 09:19:26                                  & \ \ 11:33:20 \\
 Dec.\ (J2000.0)       &  $-$04:14:34                                 &  $-$11:07:22                                 &  $+$14:56:29                                 &  $+$00:50:14                                  & $+$55:23:27 \\
 $l$                   &   160\fdg17                                  & 248\fdg41                                    &  \,75\fdg12                                  & 231\fdg06                                     & 144\fdg17 \\
 $b$                   & $-$62\fdg27                                  & +32\fdg30                                    & $-$32\fdg60                                  & +32\fdg83                                     & 58\fdg40 \\
 (m$-$M)$_0$           & 14.5~$\pm$~0.5                               & 15.8~$\pm$~0.5                               & 16.2~$\pm$~0.5                               & 16.8~$\pm$~0.5                                & 15.5~$\pm$~0.5 \\
 Median E(B$-$V)       & 0.03                                         & 0.06                                         & 0.10                                         & 0.05                                          & 0.02     \\
 Heliocentric distance & 7.9$^{+2.1}_{-1.6}$~kpc                      & 14.5$^{+3.7}_{-3.0}$~kpc                     & 17.4$^{+4.5}_{-3.6}$~kpc                     & 22.9$^{+5.9}_{-4.7}$~kpc                      & 12.6$^{+3.3}_{-2.6}$~kpc \\
 Width (FWHM)          & 27$\arcmin\,\pm$\,3$\arcmin$\,(63\,$\pm$\,7\,pc) & 27$\arcmin\,\pm$\,3$\arcmin$\,(112\,$\pm$\,14\,pc) & 20$\arcmin\,\pm$\,4$\arcmin$\,(99\,$\pm$\,20\,pc) & 52$\arcmin\,\pm$\,6$\arcmin$\,(350\,$\pm$\,40\,pc) & 37$\arcmin\,\pm$~6$\arcmin$\,(140~$\pm$~20~pc) \\
 Length                & $\sim 5\degr$ ($\sim 700$~pc)                & $\sim 10\degr$ ($\sim 2.5$~kpc)              & $\sim 8\degr$ ($\sim 2.4$~kpc)               & $\sim 45\degr$ ($\sim 21$~kpc)                & $\sim 25\degr$ ($\sim 5.5$~kpc) \\
 $M_V$                 & $-1.0^{+0.5}_{-0.7}$                         & $-2.8^{+0.3}_{-0.5}$                         & $-2.7^{+0.4}_{-0.6}$                         & $-4.9^{+0.2}_{-0.3}$                          & $-1.9^{+0.4}_{-0.6}$ \\
 $L_V$                 & $0.22^{+0.19}_{-0.08} \times 10^3 L_{\sun}$  & $1.2^{+0.8}_{-0.3} \times 10^3 L_{\sun}$     & $1.0^{+0.7}_{-0.3} \times 10^3 L_{\sun}$     & $7.5^{+2.0}_{-1.3} \times 10^3 L_{\sun}$      & $0.50^{+0.40}_{-0.15} \times 10^3 L_{\sun}$ \\
 Pole ($\alpha$, $\delta$) & (300\fdg856, 20\fdg732)                  & (65\fdg603, 32\fdg567)                       & (232\fdg227, 33\fdg838)                      & (49\fdg640, 2\fdg467)                       & (42\fdg526, 23\fdg987) \\
 Detection significance & 7.8-$\sigma$ & 6.6-$\sigma$ & 4.6-$\sigma$ & 5.5-$\sigma$ & 4.3-$\sigma$ \\
\hline
\end{tabular}
\end{table*}


There is also a broad component at ($\alpha,\delta$)$\sim$(80$\degr$, 0$\degr$), running parallel to the Milky Way disc, labelled `Mon?' in the middle panel. In the top panel it appears to be part of the disc flaring (although with a sharper southwest edge), but it is still clearly visible in the other two panels where the disc flaring feature has faded. We have checked that the appearance of this feature is not an effect of the reddening in this part of the sky by comparison with the dust maps from \citet{sch14}.
It is intriguing that Fig.\ 13 of \citet{mor16}, which presents the heliocentric distance to the Monoceros Ring centre of mass along each line of sight, shows a clear feature with a similar shape at the same location; it is distinguished by being located roughly 5~kpc further away than the surrounding stars.
Further observations will be required before it can be established if this feature is merely part of the perturbed Milky Way outer disc or an accreted component.

\subsection{The Orphan stream}

The Orphan stream, so named for its lack of an obvious progenitor, was independently discovered in SDSS survey data by \citet{bel06b} and \citet{gri06c}, who mapped it over more than 60$\degr$ on the sky. With a width of $2\degr$ and a significant internal metallicity dispersion, it is most likely the result of the tidal disruption of a dwarf galaxy.
\citet{gri15} recently used the DECam on the Blanco telescope to trace the Orphan stream beyond the southern edge of the SDSS footprint. In addition to mapping the stream a further $\sim$$50\degr$, they find a
moderate overdensity of stars at $\delta\sim-14\degr$ that they suggest could be consistent with the progenitor remnant. Our maps do not reveal any clear enhancement of star counts at this location, although this part of the sky suffered from non-optimal observing conditions in the PS1 dataset. The only significant overdensity we can detect in the stream is located at ($\alpha,\delta$)~$\sim$~(11$^h$16$^m$, $-22\degr48^m$). This is also visible on the map of \citet{gri15}, although with a lower significance. A visual inspection of the stacked images did not reveal any obvious overdensity of sources at this location.

\section{New candidate streams}\label{sec:new}

Our success in recovering almost all known streams within 35~kpc that fall in the PS1 3$\pi$ Survey footprint demonstrates the high quality of the photometric catalogue and it is therefore natural to conduct a search for additional, previously-unknown substructures in this area. Based on visual inspection of our maps, we have identified several other stream candidates at various distances; the five most significant detections, for which we could make convincing plots, are labelled PS1-A to E in Figure~\ref{fig:maps}. We also show cropped maps, reprojected in the coordinate system of the streams, in the left hand panels of Figure~\ref{fig:newst}. The corresponding CMDs made by selecting stars in a narrow box running the whole length of the stream, corrected for extinction and fore- and background contamination, are shown in Figure~\ref{fig:cmds}.
The fiducial of Milky Way globular cluster M92 (NGC\,6341; [Fe/H]=-2.3) from \citet{ber14b} provides a good match to the observed MSTO of the candidate streams and is overplotted, while we have used the fiducial of M3 (NGC\,5272; [Fe/H]=-1.5) for Balbinot\,1 in panel~C to match the estimated metallicity of this cluster \citep{bal13}. Note that the distance used to shift the fiducials was obtained from the matched-filter slice in which the stream signal was strongest, rather than from a fit to the CMD features. Because of the crudeness of the method used to estimate the distance and the considerable uncertainties on the age and metallicity of the stream stellar populations, we adopt a distance uncertainty of 0.5~mag.

We assessed the significance of the stream detections by using the stellar density maps, rather than the matched-filter maps in which noise can be misleading due to the amplification of stars in certain CMD positions. We began by constructing, for each stream, a colour--magnitude selection box based on the shape of the fiducial and taking into account the photometric uncertainties as a function of magnitude, as shown in the top left panel of Figure~\ref{fig:cmds}. The stars in this box were then used to create a spatial density map with a pixel scale of 6$\arcmin$, smoothed with a Gaussian filter of FWHM = 2.5 pixels, which was then projected onto the latitude axis to produce a cross-section of the stream. These are shown in the right hand panels of Figure~\ref{fig:newst}. The significance, defined as the peak detection, is typically 4- to 8-$\sigma$ above the background noise level. The blue line is a Gaussian fit to the overdensity, from which we estimate the width and luminosity of each stream (see below). We discuss these in more detail below, and summarize the stream properties in Table~\ref{tab:prop}.

We follow a method similar to that presented in \citet{ber14a} to estimate the total luminosity of the streams.
Firstly, {\sc IAC-Star} \citep{apa04} is used to generate the CMD of an OMP population (11.5--12.5~Gyr, [Fe/H]=$-$2.2) with the Padova library \citep{gir00}, adopting a binary fraction of 15\%, typical of the observed fraction in globular clusters \citep[e.g.][]{sol07}. The CMD contains 10$^6$ stars down to M$_V=7$ (i.e.\ $\sim$3.5~mag below the MSTO) -- stars fainter than this limit have a negligible contribution to the total magnitude. The Gaussian fits to the cross-sections described above provide the excess number $N$ of stars within the colour--magnitude selection box over the background level.
We then extract stars randomly from the synthetic CMD until the selection box contains $N\pm\sigma_N$ stars, and sum their luminosity to obtain the total flux. We repeated this step 10$^4$ times to take into account the effect of stochastic sampling of the CMDs; the total magnitude and luminosity of each stream are listed in Table~\ref{tab:prop}.

\subsection{PS1-A}

PS1-A appears as a prominent, elongated overdensity in the top panel of Figure~\ref{fig:maps}, projected on the southern extension of the Sagittarius bright stream. It passes within 1$\degr$ of the Milky Way globular cluster Whiting\,1, but is four times nearer \citep[7.9 vs.\ 31.6~kpc; e.g.][]{car07} and therefore unrelated. Thanks to its proximity, it also has the most significant detection signal at 7.8-$\sigma$.
From the matched-filter map shown in Figure~\ref{fig:newst}, we estimate a projected length of $\sim$5$\degr$, while the Gaussian fit to the cross-section profile indicates a width of 27$\arcmin$. These correspond to 700 and 63~pc, respectively, at the distance of the stream. This width is comparable to the values measured for other streams resulting from the disruption of globular clusters \citep[$\sim$100~pc; see e.g.][]{gri09,kop10}. No obvious progenitor is visible on the maps.

\subsection{PS1-B}

PS1-B was found in the intermediate distance map, near the eastern edge of the EBS, where we are able to trace it over $\sim$$10\degr$. However, as shown in the middle panel of Fig.~\ref{fig:maps}, it lies very close to the expected extension of the Lethe stream \citep{gri09}; the heliocentric distances and width -- $\sim$14.5~kpc and 112~pc vs. 13~kpc and 95~pc for Lethe \citep{gri09} -- are also in good agreement given our uncertainties. If these two features are indeed linked, this would constitute one of the longest globular cluster streams known ($\sim120\degr$).

\begin{figure*}
  \includegraphics[height=9.5cm]{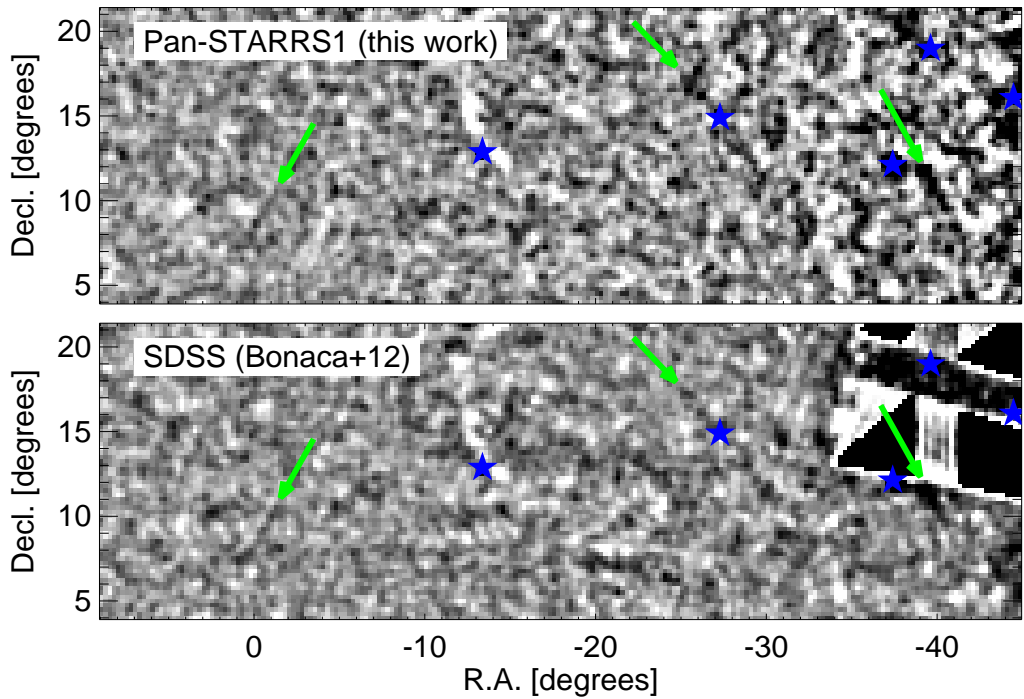}
  \caption{Comparison of our matched-filtered stellar density map at $\sim$15~kpc with that obtained by \citet{bon12} from SDSS data, centred on a small patch of the south galactic cap. Blue stars show the location of known globular clusters, while the arrows point toward some of the stream-like features that are common in both maps. The width of the smoothing kernel has been chosen to highlight very narrow stream-like features; the arrow at R.A.~$\sim-24\degr$ points to PS1-C, which is more prominent with a broader smoothing kernel.}
  \label{fig:p_vs_s}
\end{figure*}

\subsection{PS1-C}

PS1-C stretches for $\sim$$8\degr$ across the southern Galactic cap. With a width of 99$\pm$20~pc, it is consistent with a globular cluster progenitor. Interestingly, it is roughly centred on the recently discovered globular cluster Balbinot\,1 \citep{bal13}, suggesting it could represent tidal tails from this faint (M$_V=-1.21\pm0.66$), extended cluster. However, the approximate distance we estimate for the stream, based on the matched-filter slice in which the signal was strongest, is approximately 15~kpc, while the cluster lies at a distance of $31.9^{+1.0}_{-1.6}$~kpc \citep{bal13}.
This implies that either the two features are unrelated, or that our distance is strongly under-estimated.  Unfortunately, the CMD features (Figure~\ref{fig:cmds}) are not prominent enough to refine our estimate through isochrone fitting hence further observations are required.
Note, however, that this stream is also visible as a low significance overdensity in the maps of \citet{bon12} based on SDSS data (see section~\ref{sec:disc}), giving further credence to its reality.
If the stream can be proven to be physically associated with Balbinot\,1, it would contain roughly four times the luminosity of cluster, suggesting we are witnessing the object in the final throes of tidal disruption.

\subsection{PS1-D}

Compared to the other newly detected streams, PS1-D is significantly longer, broader, more luminous, and further away. In our maps, we trace it over 45$\degr$ in projection, at a distance of about 23~kpc. At this distance, the stream is 21~kpc long and 350~pc wide, i.e.\ a factor $\sim$3 broader than the other candidates described here. This suggests that the progenitor was a low-luminosity dwarf galaxy rather than a globular cluster (e.g.\ GD-1: $\sim$80~pc, \citealt{kop10} vs. Orphan: $\sim$650~pc, \citealt{bel07b}).
The CMD shown in Figure~\ref{fig:cmds} is complicated by the presence of other substructures along the same line of sight, namely the EBS, which is located at a heliocentric distance of about 10~kpc, hence the offset of $\sim$1.8~mag.

\subsection{PS1-E}

PS1-E is the most diffuse of the streams presented here, and also has the lowest significance at 4.3-$\sigma$. However, its CMD is very clean, with a well defined MSTO. As for PS1-A--C, the width of 140$\pm$20~pc suggests a globular cluster origin.
It runs over 25$\degr$-long, just a few degrees north of GD-1, which is also visible in Figure~\ref{fig:newst}. In this plot, the overdensity at ($\Lambda$,$B$)$\sim$(27.5,0.4) appears to be unrelated as it is offset by $\sim$0.5$\degr$ from the stream path, so this region has been masked before calculating the stream significance and its profile. Finally, we note that the densest parts of PS1-E are also visible in the maps of \citet{bon12}, which strongly suggests that this feature, while having a low significance, is not an artefact.

\section{Discussion and Conclusions}\label{sec:disc}

We have presented a synoptic map of Galactic halo substructures in the sky north of $\delta\sim-30\degr$ through applying the matched-filtering technique to the extant PS1 3$\upi$ Survey dataset. Covering roughly 30,000 square degrees, this is the largest deep contiguous view of the Milky Way halo yet constructed.
We have recovered almost all the previously-known stellar streams and other substructures within the volume to which we are sensitive, demonstrating the high quality and uniformity of the PS1 photometry. In addition, we have also uncovered five new candidate halo streams, one or two of which may be possible extensions of known streams and objects. Four of these streams have properties consistent with disrupting or disrupted globular clusters, while the fifth likely originates from an accreted dwarf galaxy.
Three of the globular cluster streams are short in projection, subtending $\la10\degr$ on the sky. Prior to this work, the only short GC stream known was Ophiuchus, subtending a mere $2.5\degr$ on the sky. Subsequent work has shown that Ophiuchus is likely highly foreshortened due to its inclination with our line-of-sight, but is still only $\sim$1.6~kpc after deprojection \citep{ses15}. The fact we have uncovered three similar examples of short streams in our full 3$\upi$ analysis suggests that such features may not be that rare.

Perhaps surprisingly, four of the five new streams are located within the SDSS footprint, an area that has been thoroughly searched by several groups using similar techniques to our own \citep[e.g.][]{new02,ode03,gri06b,bel07b,bon12}. In fact, PS1-A, C, and E, and the northern end of PS1-D, are in hindsight discernable as extended overdensities in the SDSS maps of \citet{bon12}\footnote{available at\\ http://www.astro.yale.edu/abonaca/research/halo.html}, although with a much lower significance than the PscTri stream that is the subject of their paper. We therefore explored the possibility of finding other low significance streams by blinking our PS1 maps with those of \citet{bon12}, who used comparable distance bins. Remarkably, this process reveals many narrow, stream-like features in common, sometimes extending over tens of degrees. It also reveals compact overdensities which do not correspond to known globular clusters or dwarf galaxies, although many of those may be due to the presence of background galaxy clusters.
To illustrate the power of this comparison, Figure~\ref{fig:p_vs_s} shows a small area of the southern galactic cap from both surveys, where extended features in common are highlighted with green arrows. Since the observing strategies, data analysis pipelines, and reddening maps used are all different and independent, the coincidence of these features indicates that they are most likely real. However, we have checked that even the most significant of these features are too faint and/or diffuse to produce convincing CMDs, and therefore we are unable to determine their physical properties with any of the extant data.

That the Milky Way halo may be composed of a myriad of faint tidal streams, most of which lie just below the detection limits of current photometric surveys, is a tantalizing prospect that has also been recently hinted at from spectroscopy \citep[e.g.][and references therein]{sch12}. If confirmed, this would give strong support for the hierarchical model of structure formation on the scale of individual galaxies.  Understanding the origins of halo streams is of tantamount importance. Many of the features identified so far are narrow and composed of ancient metal-poor populations, properties that are most consistent with disrupted globular clusters.
This suggests that a non-negligible fraction of the stellar halo may have originated in globular clusters, in agreement with chemical tagging analyses \citep[e.g.][]{mar11,mar16,ram12,lin15,fer16}. The next decade is likely to be pivotal for disentangling the actual make-up and assembly history of the Milky Way halo, with forthcoming wide-field surveys such as the \emph{Large Synoptic Survey Telescope} \citep{tys02} reaching several magnitudes beyond what is currently available. In addition, the soon availability of \emph{Gaia} astrometric data will provide crucial constraints on the distances and orbits of nearby streams, as well as facilitate the discovery of new substructures through joint photometric and kinematic searches.

\section*{Acknowledgements}

The authors are grateful to Ana Bonaca for granting us permission to use part of her maps in this publication, and to Jorge Pe\~narrubia and the anonymous referee for useful comments.
EJB acknowledges support from a consolidated grant from STFC and from the CNES postdoctoral fellowship program. HWR acknowledges support from the DFG grant SFB 881 (A3). RFGW acknowledges support through a Visiting Professorship from the Leverhulme Trust, held at the University of Edinburgh.
This research work has made use of the Python packages Numpy\footnote{http://www.numpy.org/} \citep{wal11},
Astropy\footnote{http://www.astropy.org} \citep{ast13}, Matplotlib\footnote{http://matplotlib.org/} \citep{hun07}, and Pandas\footnote{http://pandas.pydata.org/} \citep{mck10}; the IAC-STAR Synthetic CMD computation code, which is supported and maintained by the computer division of the Instituto de Astrof\'isica de Canarias; and the NASA/IPAC Extragalactic Database which is operated by the Jet Propulsion Laboratory, California Institute of Technology, under contract with the National Aeronautics and Space Administration.

The PS1 Surveys have been made possible through contributions of
the Institute for Astronomy, the University of Hawaii, the Pan-STARRS Project
Office, the Max-Planck Society and its participating institutes, the Max Planck
Institute for Astronomy, Heidelberg and the Max Planck Institute for
Extraterrestrial Physics, Garching, The Johns Hopkins University, Durham
University, the University of Edinburgh, Queen's University Belfast, the
Harvard-Smithsonian Center for Astrophysics, the Las Cumbres Observatory Global
Telescope Network Incorporated, the National Central University of Taiwan, the
Space Telescope Science Institute, the National Aeronautics and Space
Administration under Grant No.\ NNX08AR22G issued through the Planetary Science
Division of the NASA Science Mission Directorate,  the National Science
Foundation under Grant No.\ AST-1238877, and the University of Maryland.



\label{lastpage}
\end{document}